\documentclass[12pt]{article}

\usepackage{epsfig}
\usepackage{graphics}
\usepackage{multicol}

\newcommand{\fig} [1] {Fig.~\ref{#1}}

\newcommand{\tab} [1] {Table~\ref{#1}}

\def\be{\begin{equation}}
\def\ee{\end{equation}}

\begin{document}

\begin{flushright}
\small ZU-TH 19/02
\end{flushright}

\begin{center}
{\bf \Large 
Reconstruction of Supersymmetric High Scale Theories}\\[0.5cm]
{\large  W.~Porod\\[0.5cm] 
\small
 Inst.~f\"ur Theor. Physik, Universit\"at Z\"urich, CH-8057 Z\"urich,
      Switzerland 
}
\end{center}

\begin{abstract}
We demonstrate how the fundamental supersymmetric theory at high
energy scales can be reconstructed using precision data expected
at future high energy collider experiments. 
We have studied a set of representative examples in this context: 
minimal supergravity; gauge mediated supersymmetry breaking;
and superstring effective field theories.
\end{abstract}

\section{Introduction}

Supersymmetry is one of the most attractive extensions of the 
Standard Model. Therefore the discovery of supersymmetric
particles as well as the accurate measurement of their properties
are among the main topics in the experimental programs for
future high--energy colliders, such as LHC \cite{LHC} and prospective
$e^+ e^-$ linear colliders \cite{LC}. 
In this report we summarize how high--precision measurements of supersymmetric
particles can be used to extract information on the underlying
high scale theory at the GUT / Planck and to eventually reconstruct this theory
\cite{Blair:2000gy}. 

The general strategy is based on the bottom--up approach. It can be
summarized shortly as follows (see \cite{Blair:2000gy} for more details).
From a specific theory at the high scale
one calculates the observables at the electroweak scale, e.g.~masses
and cross sections. These observables are endowed with errors as
expected at future high--energy experiments. This
set of data is adopted
as input and the origin of data in terms
of the high scale theory is ``forgotten''. 
Next one extracts the parameters plus
the corresponding errors at
the electroweak scale from the experimental observables. These
parameters are extrapolated to the high scale by means of renormalization
group techniques. In this way one gains insight to which extent
and with which accuracy the orginal theory can be reconstructed. In the
following we exemplify this procedure by considering three examples: 
supergravity \cite{sugra}, gauge mediated supersymmetry breaking \cite{gmsb}, 
and superstring effective field theories motivated by orbifold compactified
heterotic string theories \cite{cvetic}. 

\section{Gravity Mediated SUSY Breaking}

Supersymmetry is broken in a hidden sector in supergravity models (SUGRA) and
the breaking is transmitted to our eigenworld by gravitational interactions
\cite{sugra}.
In this scheme it is suggestive although not compulsory that the soft SUSY
breaking parameters are universal at the high scale, e.g.~the GUT scale
$M_U$.

The SUGRA point we have analyzed in detail, was chosen close to 
the Snowmass Point SPS\#1
\cite{Allanach:2002nj}, except for the scalar mass parameter
$M_0$ which was taken slightly larger for merely
illustrative purpose: $M_{1/2} = 250$~GeV, $M_0 = 200$~GeV, $A_0 = -100$~GeV,
$\tan \beta = 10$ and $sign(\mu) = +$. 
This set of parameters is compatible with the present results of 
low--energy experiments.
The initial ``experimental'' values, 
have been generated by evolving the universal parameters down to the 
electroweak scale according to standard
procedures \cite{aarason,bagger}.
 These parameters
 define the experimental observables, including
supersymmetric particle masses and production cross sections. They are
endowed with errors as expected for 
threshold scans as well as  for measurements  
in the continuum  at $e^+ e^-$ linear colliders (LC). 
The errors given in  Ref.\cite{blair2} are scaled
in proportion to the masses of the spectrum. 
Typical examples are shown in Table~\ref{tab:masserrors}.
The LC errors on the squark masses, see {\it e.g.} Ref.\cite{Feng},
are set to an average value of 10 GeV. 
For the cross-sections we use only statistical errors, while
assuming a (conservative) reconstruction efficiency of 20\%.  
These observables are interpreted as the experimental input values for
the evolution of the mass parameters in the bottom-up
approach to the Grand Unification scale. 
\begin{table}
\begin{center}
\begin{tabular}{c|cc||c|cc}
Particle           & M(GeV) & $\Delta$ M(GeV) &
Particle           & M(GeV) & $\Delta$ M(GeV)\\\hline \hline
$h^0$              &  113.33      & 0.05  &
 $\tilde{e}_L$      & 269.1 & 0.3 \\
$A^0$              &  435.5        &  1.5  &
 $\tilde{e}_R$      & 224.82    &       0.15\\
$\tilde{\chi}^+_1$         & 183.05         & 0.15   &
$\tilde{\tau}_1$   & 217.7    &       1.00 \\
$\tilde{\chi}^+_2$         & 383.3         & 0.3   &
$\tilde{\tau}_2$   & 271.5    &       0.9 \\
$\tilde{\chi}^0_1$         & 97.86        &  0.2 &
$\tilde{u}_L$      & 589    &       10  \\
$\tilde{\chi}^0_2$         & 184.6        &  0.3 &
$\tilde{u}_R$      & 572    &       10  \\
$\tilde{g}$        & 598    &   10 &
$\tilde{t}_1$      & 412    &       10  \\ \hline
\end{tabular}
\end{center}
\caption[]{\it Representative experimental mass errors used in the
fits to the mass spectra; with the exception of the gluino mass, all
the other parameters are based on LC measurements.}
\label{tab:masserrors}
\end{table}

\begin{figure}
\setlength{\unitlength}{1mm}
\begin{center}
\begin{picture}(160,140)
\put(-4,-3){\mbox{\epsfig{figure=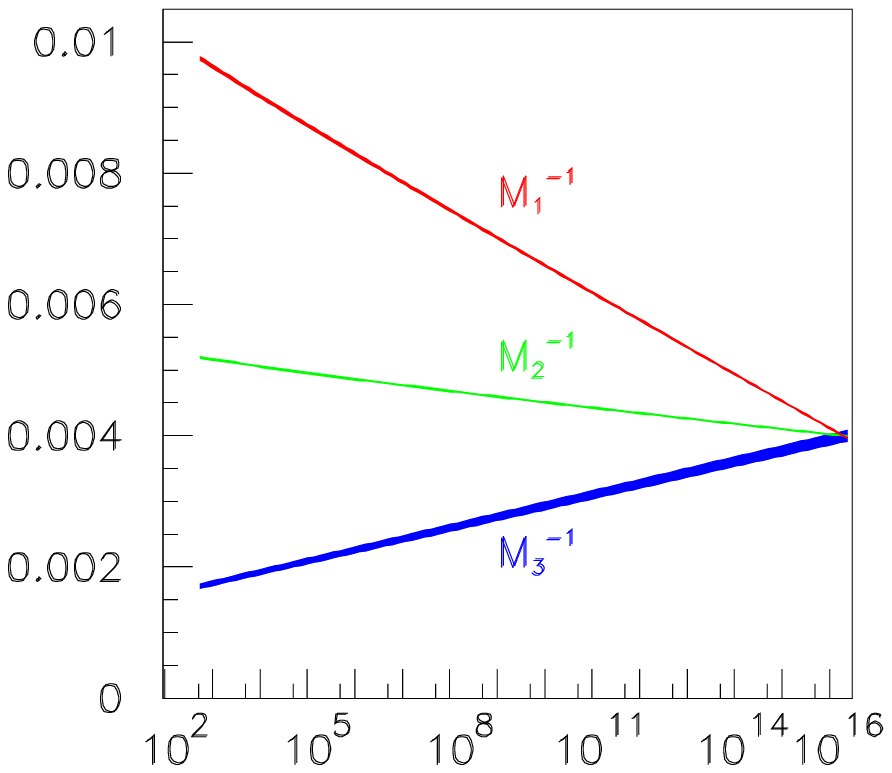,height=15cm,width=17cm}}}
\put(81,69){\mbox{\epsfig{figure=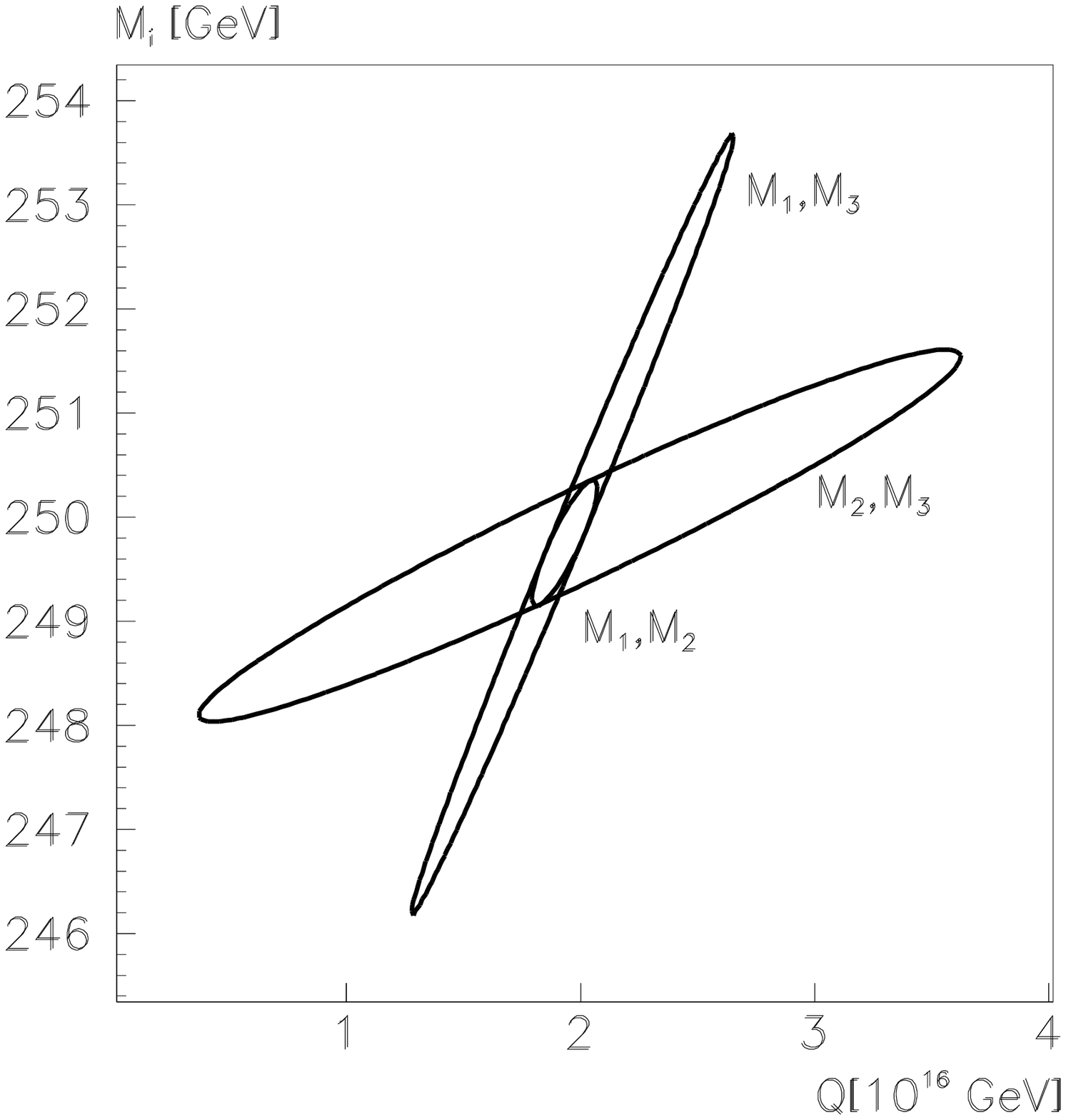,
                                   height=7.0cm,width=7.8cm}}}
\put(-4,-76){\mbox{\epsfig{figure=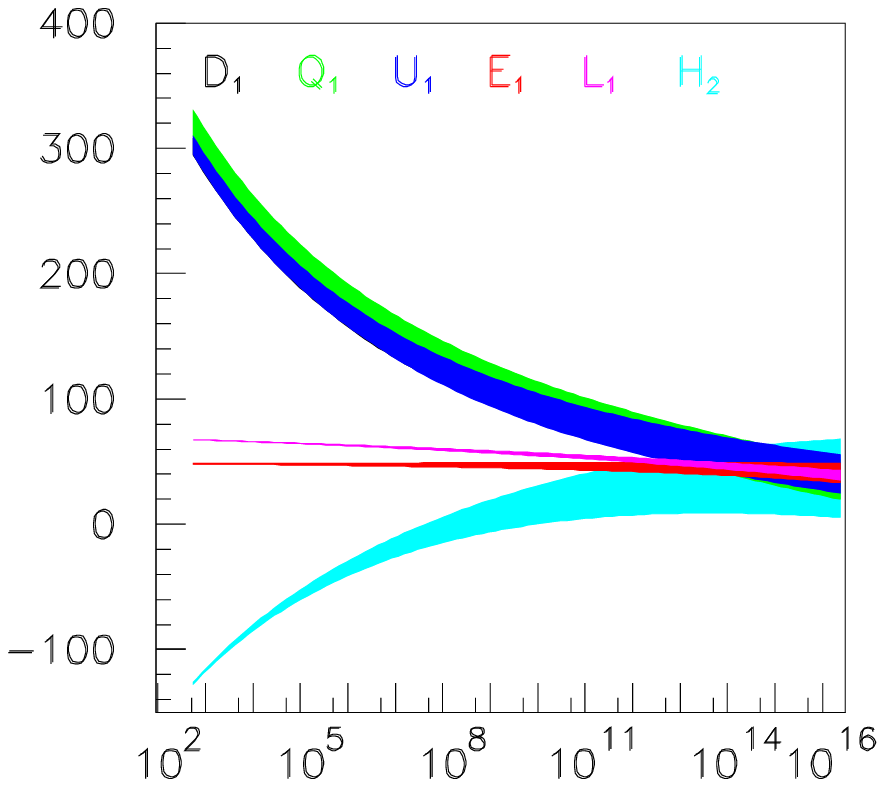,height=15cm,width=17cm}}}
\put(76,-76){\mbox{\epsfig{figure=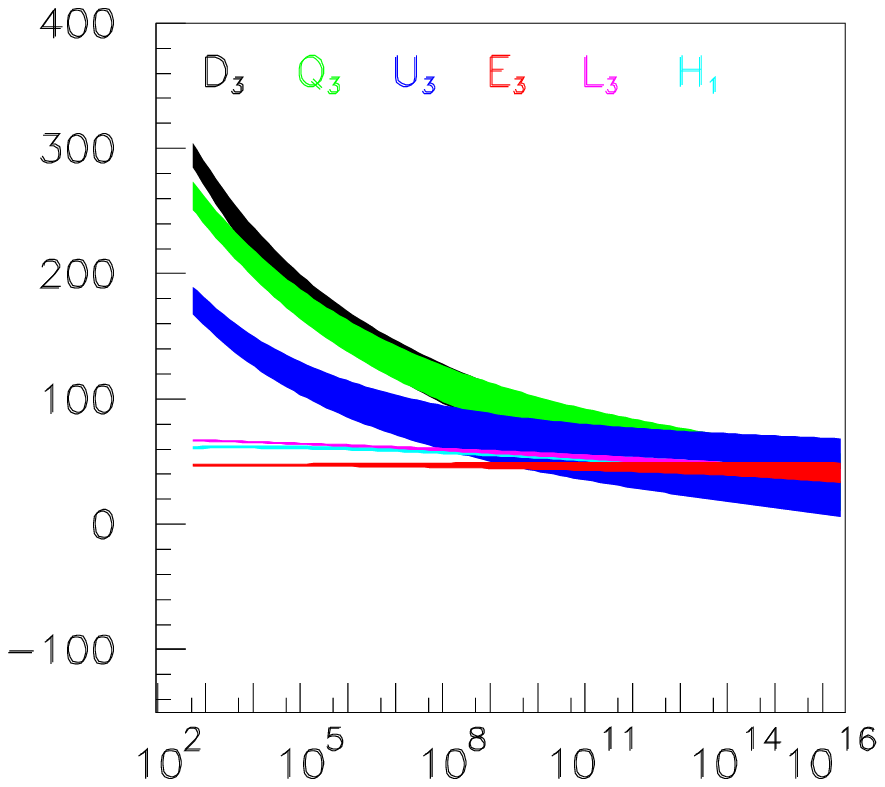,height=15cm,width=17cm}}}
\put(-1,136){\mbox{\bf (a)}}
\put(14,135){\mbox{$1/M_i$~[GeV$^{-1}$]}}
\put(62,70){\mbox{$Q$~[GeV]}}
\put(80,136){\mbox{\bf (b)}}
\put(-1,63){\mbox{\bf (c)}}
\put(14,62){\mbox{$M^2_{\tilde j}$~[$10^3$ GeV$^2$]}}
\put(63,-4){\mbox{$Q$~[GeV]}}
\put(80,62){\mbox{\bf (d)}}
\put(95,62){\mbox{$M^2_{\tilde j}$~[$10^3$ GeV$^2$]}}
\put(142,-4){\mbox{$Q$~[GeV]}}
\end{picture}
\end{center}
\caption{{\bf mSUGRA:} {\it  Evolution, from low to high scales, of 
(a) gaugino mass parameters, and (b) unification of gaugino mass parameter
   pairs; 
 (c) evolution of first--generation sfermion mass parameters and 
     the Higgs mass parameter $M^2_{H_2}$;
 (d) evolution of third--generation sfermion mass parameters and  
     the Higgs mass parameter $M^2_{H_1}$. 
 The initial
parameters are given by: $M_0 = 200$~GeV, 
$M_{1/2} = 250$~GeV, $A_0$ = -100~GeV, $\tan \beta = 10$, 
and $\mathrm{sign}(\mu) = (+)$.
[The widths of the bands indicate the 1$\sigma$ CL.]
}}
\label{fig:sugra}
\end{figure} 

The presumably strongest support, though indirect, 
for supersymmetry is related to the
tremendous success of this theory in predicting the unification of the
gauge couplings \cite{GUT}. The precision, being  at the per--cent level, is
surprisingly high after extrapolations over
fourteen orders of magnitude in the energy 
from the electroweak scale to the unification scale $M_U$. 
The expected accuracies in $M_U$ and $\alpha_U$, based on the GigaZ
option, are: $M_U = (2.000 \pm 0.016) \cdot 10^{16} \, \rm {GeV}$ 
and 
$\alpha_U^{-1}= 24.361 \pm 0.007 $.

For the evolution of the gaugino and scalar mass parameters two--loop
RGEs \cite{RGE2} have been used. 
One--loop threshold effects are incorporated using the formulas
given in \cite{bagger} and in case of Higgs bosons we have  included two--loop
effects as given in \cite{Degrassi:2001yf}.
The results for the evolution of the mass parameters to the GUT scale $M_U$
are shown in \fig{fig:sugra}.
 \fig{fig:sugra}(a)  presents the evolution of the gaugino
parameters $M_i^{-1}$ which clearly is under excellent control, 
as are the extrapolations
of the slepton mass parameters squared of the first (and second) and the
third generation
in Fig.~\ref{fig:sugra}(c) and (d), respectively. The accuracy
deteriorates for the squark mass parameters and for the Higgs mass parameter
$M^2_{H_2}$.
The origin of the differences between the errors for slepton, squark and
Higgs mass parameters can be traced back to the structure of the RGEs.
This can easily be understood by inspecting the 
approximate solutions of the RGEs.
Typical examples evaluated at the scale $Q=500$~GeV read as
follows:
\begin{eqnarray}
M^2_{\tilde L_{1}} &\simeq& M_0^{2} + 0.47 M^2_{1/2} \\
M^2_{\tilde Q_{1}} &\simeq& M_0^{2} + 5.0 M^2_{1/2}  \\
M^2_{\tilde H_2} &\simeq&  -0.03 M_0^{2} - 1.34 M^2_{1/2}
           + 1.5 A_0 M_{1/2} + 0.6 A^2_0
\end{eqnarray}
While the coefficients for sleptons are of order unity,
the coefficient for squarks in front of $M^2_{1/2}$ is 5, so that small errors
in $M^2_{1/2}$ are magnified by nearly an order of magnitude in the solution
for $M_0$. This feature becomes even more enhanced for the Higgs mass
parameter, giving rise to large errors in this case.
A representative
set of mass values and the associated errors,  
after evolution from the electroweak scale to $M_U$, is
presented in Table~\ref{tab:parvalues_a}.
The corresponding error ellipses for the unification of the 
gaugino masses are shown in
\fig{fig:sugra}(b).
\begin{table}
\begin{center}
\begin{tabular}{c||c|c}
 &  Exp.~Input &  GUT Value \\ \hline   \hline
 $M_1$~[GeV] & 102.31 $\pm$  0.25 &  $250.00 \pm  0.33$ \\
 $M_2 $~[GeV] &  192.24 $\pm$  0.48      &  $250.00 \pm  0.52$ \\
 $M_3 $~[GeV] & 586  $\pm$  12   &  $250.0    \pm   5.3$  \\ \hline
$\mu$         & 358.23  $\pm$ 0.28     &  $355.6 \pm  1.2    $  \\
\hline
 $M^2_{L_1} $~[GeV$^2$] & $( 6.768  \pm  0.005)\cdot 10^4$
                &  $(3.99  \pm  0.41) \cdot 10^4$  \\
 $M^2_{E_1} $~[GeV$^2$] & $(4.835  \pm  0.007) \cdot 10^4$
  &  $(4.02  \pm  0.82)  \cdot 10^4 $ \\
 $M^2_{Q_1} $~[GeV$^2$] &  $(3.27 \pm  0.08)\cdot 10^5$
               &  $(3.9  \pm  1.5) \cdot 10^4$ 
\\ \hline 
 $M^2_{L_3} $~[GeV$^2$] & $(6.711 \pm  0.050)\cdot 10^4$
   &  $(4.00  \pm  0.41)  \cdot 10^4 $  \\
 $M^2_{E_3} $~[GeV$^2$] & $(4.700 \pm  0.087)\cdot 10^4$
  &  $(4.03  \pm  0.83) \cdot 10^4 $  \\
 $M^2_{Q_3} $~[GeV$^2$] &  $(2.65 \pm  0.10) \cdot 10^5$
  &  $(4.1  \pm  3.0)  \cdot 10^4 $ \\
\hline
 $M^2_{H_1} $~[GeV$^2$] &  $(6.21 \pm  0.08)\cdot 10^4$  &
  $ (4.01  \pm  0.54)  \cdot 10^4 $ \\
 $M^2_{H_2}$~[GeV$^2$] &  $(-1.298 \pm 0.004)\cdot 10^5$ &
  $(4.1  \pm  3.2) \cdot 10^4 $\\
\end{tabular}
\end{center}
\caption[]{\it Representative gaugino/scalar mass parameters and couplings
as determined
at the electroweak
scale and evolved to the GUT scale in the mSUGRA scenario;
based on LHC
and LC simulations.  [The errors quoted correspond to 1$\sigma$.]}
\label{tab:parvalues_a}
\end{table}

Inspecting Figs.~\ref{fig:sugra}(c) and (d) leads us to the conclusion that a
blind top-down approach eventually may generate an incomplete picture. 
Global fits based on mSUGRA without allowing for deviations
from universality, are dominated by $M_{1,2}$ and the slepton mass
parameters due to the pseudo-fixed point behaviour of the squark mass
parameters.  Therefore, the structure of the theory in the squark sector
is not scrutinized stringently at the unification scale
in the top-down approach let alone the Higgs sector. 
By contrast, the bottom-up approach demonstrates very clearly the extent
to which the theory can be tested at the high scale quantitatively.
The quality of the global 
test is apparent from \tab{tab:parvalues_a}, in which 
the evolved gaugino
values should reproduce the universal mass $M_{1/2} = 250$~GeV and all
the scalars the mass $M_0 = 200$~GeV. They are compared with the global
mSUGRA fit of the universal parameters where we find 
$M_{1/2} = 250\pm 0.08$~GeV and $M_0 = 200\pm 0.09$~GeV.

\section{Gauge Mediated Supersymmetry Breaking}

\begin{figure}[t]
\setlength{\unitlength}{1mm}
\begin{center}
\begin{picture}(160,67)
\put(-4,-78){\mbox{\epsfig{figure=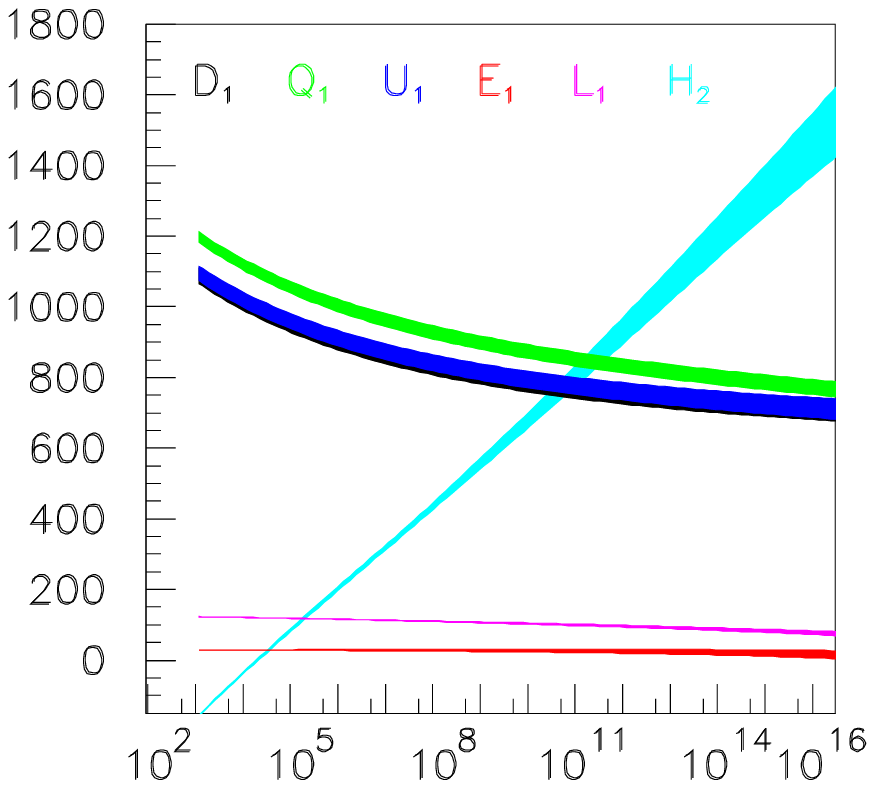,height=15cm,width=18cm}}}
\put(75,-6){\mbox{\epsfig{figure=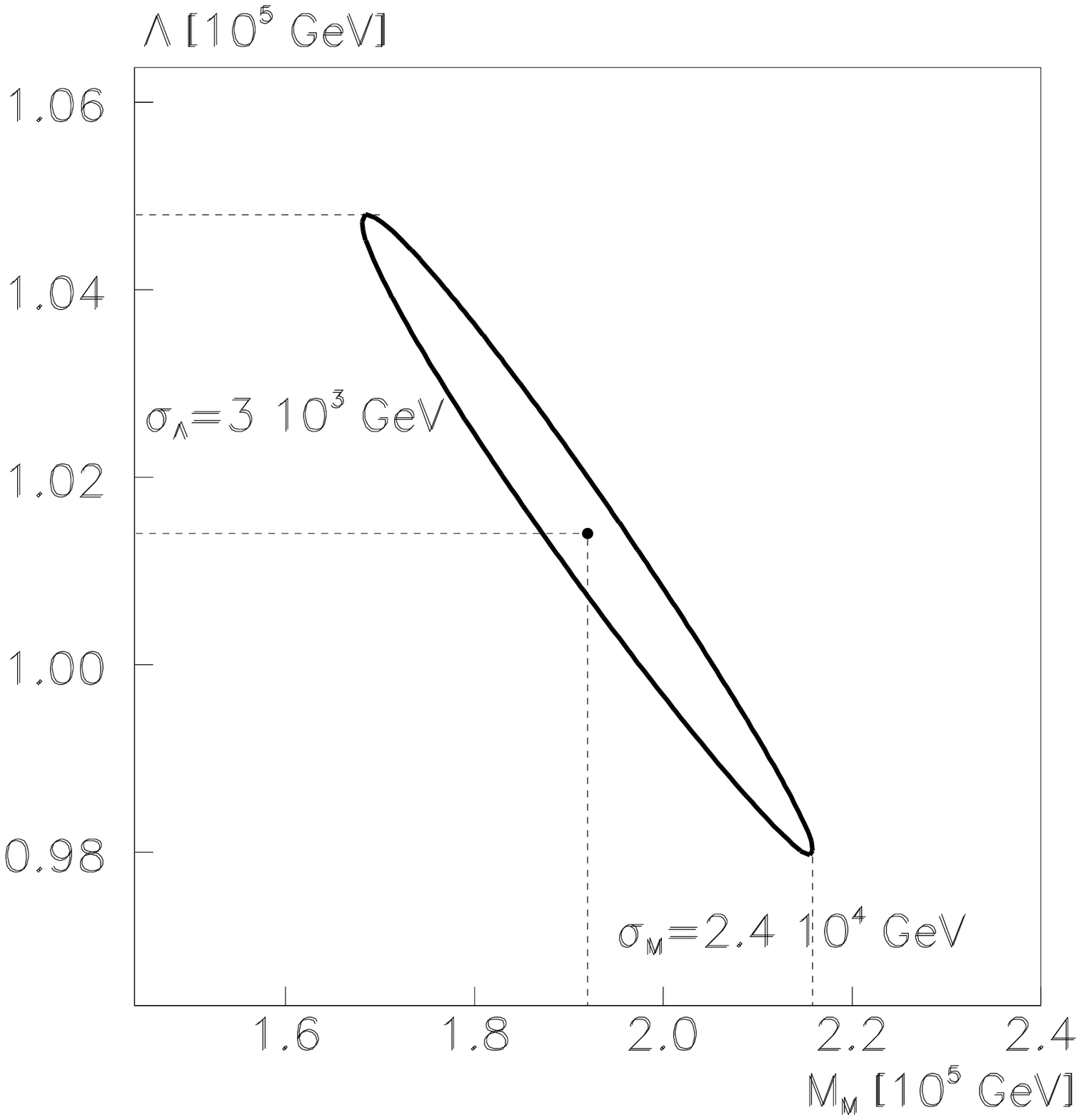,
                                   height=7cm,width=8.4cm}}}
\put(-1,62){\mbox{\bf (a)}}
\put(14,60){\mbox{$M^2_{\tilde j}$~[$10^3$ GeV$^2$]}}
\put(65,-5){\mbox{$Q$~[GeV]}}
\put(27,21.5){$M_M$}
\put(28.5,21){\vector(0,-1){5}}
\put(80,62){\mbox{\bf (b)}}
\end{picture}
\end{center}
\caption{{\bf GMSB:} {\it  Evolution of 
(a) first--generation sfermion mass parameters and 
     Higgs mass parameter $M^2_{H_2}$ and
 (b) $\Lambda$--$M_M$ determination
 in the
bottom--up approach.
The point probed, SPS\#8, is characterized by the
parameters $M_M = 200$~TeV, 
$\Lambda = 100$~TeV, $N_5 = 1$, $\tan \beta = 15$, 
and $\mathrm{sign}(\mu) = (+)$.
[The widths of the bands indicate the 1$\sigma$ CL.]
}}
\label{fig:Gmsb}
\end{figure} 

In gauge mediated supersymmetry breaking (GMSB) \cite{gmsb}
the scalar and the F components of a Standard--Model singlet superfield $S$
acquire vacuum expectation values $\langle S \rangle$ and 
$\langle F_S \rangle $ through interactions with other
fields in the secluded sector, thus breaking supersymmetry. 
The messenger fields mediating
the breaking to our eigen-world and the two
vacuum expectation values characterize
the system. The general scale is given by the messenger mass
$M_M \sim$ $\langle S \rangle $ whereas
the size of  gaugino and scalar masses is set by 
$\Lambda = {\langle F_S \rangle} / {\langle S \rangle} $.
The gaugino masses are generated at 1--loop level 
by loops of scalar and fermionic  messenger component
fields. Masses of the scalar fields in the
visible sector
are generated by 2-loop effects of gauge/gaugino and messenger fields.
The masses are equal  at the messenger scale $M_M$
for scalar particles with identical Standard--Model 
charges squared.
In the minimal version of GMSB, the $A$ parameters
are generated at 3-loop level and they are practically zero at $M_M$.

We have investigated this scheme for the point $\Lambda = 100$~TeV,
$M_M = 200$~TeV, $N_5=1$, $N_{10}=0$, $\tan\beta=15$ and $\mu>0$
corresponding to the Snowmass Point SPS\#8 \cite{Allanach:2002nj}. 
The evolution of the sfermion mass parameters
of the first generation as well as the Higgs mass parameter $m^2_{H_2}$
is shown in \fig{fig:Gmsb}a. 
It is obvious from the figure that the GMSB scenario cannot be confused
with the universal supergravity scenario. [Specific experimental
signatures generated in the decays of the next to lightest supersymmetric
particle
to gravitinos provide a complementary 
experimental discriminant, see e.g.~\cite{blair}].

The bands of the slepton $L$--doublet mass parameter $M^2_{\tilde L}$ and the
Higgs parameter $M^2_{H_2}$, which carry the same moduli of
standard--model charges, cross at the scale $M_M$. The crossing,
which is indicated by an arrow in \fig{fig:Gmsb}(a), is a necessary
condition for the GMSB scenario to be realized. 
The determination of scalar and gaugino mass parameters at this
``meeting point'' can be used to extract $\Lambda$, $M_M$ and
the multiplicity coefficient $(N_5 + 3 N_{10})$.
  For the point
analyzed one finds:
\begin{eqnarray}
\Lambda &=& (1.01 \pm 0.03) \cdot 10^2 \; \rm {TeV}\\
M_M &=&(1.92 \pm 0.24) \cdot 10^2 \; \rm {TeV} \\
N_5 + 3 N_{10} &=& 0.978 \pm 0.056
\end{eqnarray}
in good agreement with the theoretical ideal input values. 
The correlation between $\Lambda$ and $M_M$ is shown in \fig{fig:Gmsb}(b).

\section{String Induced Supersymmetry Breaking}

Superstring theories are among the most exciting candidates for a
comprehensive theory of matter and interactions. Here we summarize
results obtained for a string effective theory in four dimensions
based on orbifold compactification of the heterotic superstring
\cite{cvetic}. SUSY breaking is generated by non--pertubative effects,
mediated by a Goldstino field which is a superposition of the dilaton
field $S$ and the moduli field $T$ [all moduli fields are assumed to
be identical]: 
$\tilde G = \sin \theta \, \tilde{S} + \cos \theta \, \tilde{T}$.
 Universality is generally broken in such a scenario by
a set of non--universal modular weights $\{n_j\}$
that determine the couplings of
$T$ to the SUSY matter fields $\Phi_j$.
\begin{eqnarray}
M_i &=&  - g_i^2 m_{3/2} s {\sqrt{3} \sin \theta} + \dots \\
M_{\tilde j}^2 &=& m^2_{3/2} \left( 1 + n_j \cos^2 \theta \right) + \dots 
\end{eqnarray}
In next--to--leading order, indicated by the ellipses, also the vacuum
value $<T>$ and the Green--Schwarz parameter $\delta_{GS}$ enter.
The one loop effects give rise to non--universal corrections for the 
gaugino mass parameters as well as for
 the  gauge couplings at the unification scale $M_{GUT}$.
A precise measurement sensitive to these one--loop effects for
 the gauge couplings and the SUSY parameters
can thus be used to get information about the string scenario. 

In \fig{fig:String} we show the evolution of the gaugino mass parameters
with the crucial high--scale region expanded in the insert. Here
we have taken the parameters presented in  \tab{tab:parameters_string} as
input. Relevant
parameters derived from an overall--fit to couplings and masses are
given in \tab{tab:parameters_string}. One clearly sees that the ideal
data, from which the experimental observables were deduced, can indeed
by extracted from the data collected at future hadron-- and lepton colliders
performing high precision measurements. 

\newpage
\begin{multicols}{2}
\setlength{\unitlength}{1mm}
\begin{picture}(70,75)
\put(-3,2){\mbox{\epsfig{figure=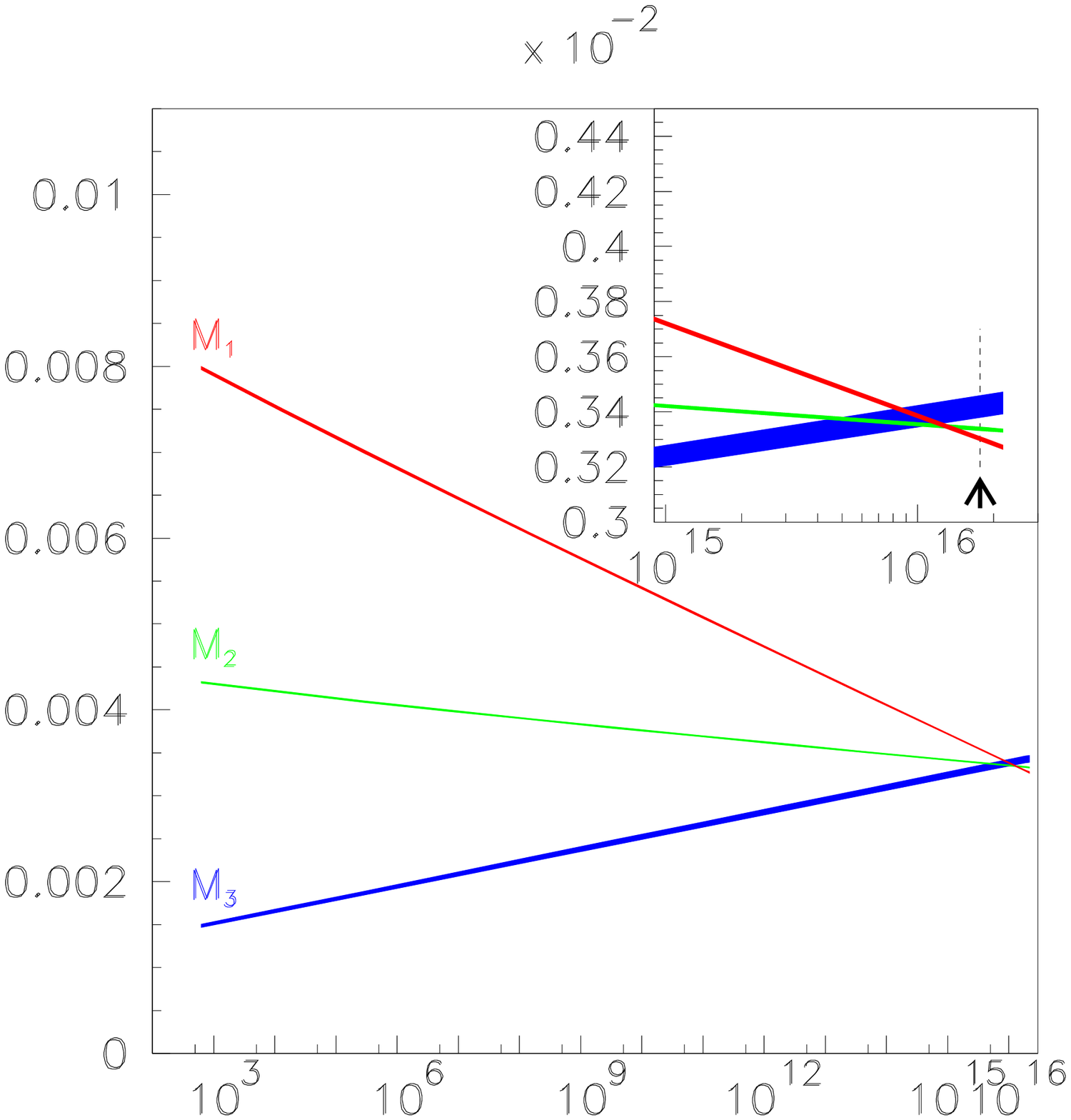,
                                   height=7.cm,width=7.5cm}}}
\put(0,68){\mbox{\small $-1/M_i$~[GeV$^{-1}$]}}
\put(55,0){\mbox{\small $Q$~[GeV]}}
\end{picture}
Figure 3: { {\bf String Scenario:} \it  Evolution of 
  gaugino mass parameters [the insert expands on the breaking
     of universality at the GUT scale].
 }
\refstepcounter{figure}
\label{fig:String}
\\

%
\begin{tabular}{c||c|rcl}
Parameter           & Ideal & \multicolumn{3}{c}{Reconstructed} \\ \hline\hline
$m_{3/2}$           &  180  &     179.9 & $\pm$ & 0.4 \\
$\langle S \rangle$ &   2   &      1.998 & $\pm$ & 0.006 \\
$\langle T \rangle$ &  14   &      14.6 & $\pm$ & 0.2 \\
$\sin^2\theta$        & 0.9 &      0.899 & $\pm$ & 0.002 \\
$\delta_{GS}$       &   0   &      0.1 & $\pm$ & 0.4 \\ \hline
$n_L$               &  -3   &      -2.94 & $\pm$ & 0.04 \\
$n_E$               &  -1   &     -1.00 & $\pm$ & 0.05 \\
$n_Q$               &   0   &     0.02 & $\pm$ & 0.02 \\
$n_U$               &  -2   &     -2.01 & $\pm$ & 0.02 \\
$n_D$               &  +1   &      0.80 & $\pm$ & 0.04 \\
$n_{H_1}$           &  -1   &      -0.96 & $\pm$ & 0.06 \\
$n_{H_2}$           &  -1   &      -1.00 & $\pm$ & 0.02 \\ \hline
$\tan \beta$        &  10   &      10.00 & $\pm$ & 0.13 \\ 
\end{tabular}
Table~3: {\it Comparison of the experimentally reconstructed values with the
               ideal fundamental parameters in a specific example for a string 
               effective field theory.} 
\refstepcounter{table}
\label{tab:parameters_string}

\end{multicols}

\section{Conclusions}

In this report we have demonstrated that fundamental parameters
of the underlying supersymmetric theory at the high scale can
be reconstructed in practice. The reconstruction is based on future
high--precision data from $e^+ e^-$ linear colliders, TESLA in particular,
combined with results from LHC and CLIC.
The bottom--up approach of evolving the parameters from the electroweak
scale to the high scale provides a transparent picture of the
underlying theory at the high scale. 
We have exemplified this conclusions  in the cases of minimal supergravity
theories, gauge mediated supersymmetry breaking, and for a string
effective theory. In the latter example we have demonstrated that one can
 indeed extract the string parameters 
-- a truly exciting observation -- from future high--precision
measurements at hadron-- and lepton--colliders.

\section*{Acknowledgements}

I would like to thank 
G.A.~Blair and P.M.~Zerwas for a very interesting and
fruitful collaboration.
This work is supported by the Erwin
Schr\"odinger fellowship No. J2095 of the `Fonds zur
F\"orderung der wissenschaftlichen Forschung' of Austria FWF and
partly by the Swiss `Nationalfonds'.

\end{document}